# Estimation and Prediction of technical debt: a proposal

Alvine Boaye Belle

Thompson Rivers University, 805 TRU Way, Kamloops BC V2C 0C8, Canada

aboayebelle@tru.ca

## Abstract

Technical debt is a metaphor used to convey the idea that doing things in a "quick and dirty" way when designing and constructing a software leads to a situation where one incurs more and more deferred future expenses. Similarly to financial debt, technical debt requires payment of interest in the form of the additional development effort that could have been avoided if the quick and dirty design choices have not been made.

Technical debt applies to all the aspects of software development, spanning from initial requirements analysis to deployment, and software evolution.

Technical debt is becoming very popular from scientific and industrial perspectives. In particular, there is an increase in the number of related papers over the years. There is also an increase in the number of related tools and of their adoption in the industry, especially since technical debt is very pricey and therefore needs to be managed.

However, techniques to estimate technical debt are inadequate, insufficient since they mostly focus on requirements, code, and test, disregarding key artifacts such as the software architecture and the technologies used by the software at hand. Besides, despite its high relevance, technical debt prediction is one of the least explored aspects of technical debt.

To address these shortcomings, it is mandatory that I undertake research to: 1) improve existing techniques to properly estimate technical debt; 2) to determine the extent to which the use of prediction techniques to foresee and therefore avoid technical debt could help companies save money and avoid a potential bankruptcy in the subsequent years.

The proposed research can have an important economic impact by helping companies save several millions. It can have a major scientific impact by leading to key findings that will be disseminated through patents, well-established scientific journals and conferences.

## Proposed Research Project

**Proposed research** – Deferring or neglecting maintenance activities aiming at fixing inefficiencies introduced in the production code has an adverse impact on many design-time qualities, schedule and costs and incurs a financial overhead due to a quality decay (Ampatzoglou, 2015). That financial overhead is known as technical debt. Introduced in the early nineties by Ward Cunningham (Cunningham, 1993), the concept of technical debt refers to the financial consequences of a "quick and dirty design" choice that is required to make trade-offs between the reduction of software time to market and the poor specification/ implementation of a software, through its development phases (Avgeriou et al., 2016).

Similarly to financial debt, technical debt requires payment of interest in the form of the additional effort that could have been avoided if the quick and dirty design choices have not been made (Kruchten et al., 2013). Technical debt applies to every single activity of the software lifecycle (Kruchten et al., 2012; Ampatzoglou, 2015). It therefore applies to all the aspects of software development, spanning from deployment, selling, or even software evolution (Falessi and Kruchten, 2015). Hence, it covers various topics such as software architecture, software quality, software maintenance or even code smells. This leads to different dimensions of technical debt. These include: architecture debt, build debt, requirement debt, code debt, documentation debt, technology debt, and test debt (Tom et al., 2013; Martini and Bosch, 2017; Alves et al., 2014). For instance, architecture debt is due to existing sub-optimal architectural solutions (e.g., existence of structural violations, components lacking modularity, architectural smells), and is the most challenging to uncover due to the lack of tools to identify and manage it. Code debt is caused by every hack that makes the code brittle and will make it break when modified in the future (e.g., duplicated and complex code, bad coding style that decreases the code readability, code that lacks logic).

As technical debt accrues, it drives a company to suffer from delays, feature quality loss, and struggles in maintaining its system's operations (Seaman et al., 2012). Two alternatives arise from this situation (Fowler, 2019). First, keep paying the interest on the debt. Second, pay down the principal of the debt by restructuring the quick and dirty design into a better one. This second alternative is preferable on a long-term basis since it avoids compounding of the debt interest (Fowler, 2019).

Technical debt has recently gained a huge popularity in industry as well as in the research community. As a consequence, there is an increase of the number of related tools and of their adoption in the industry. Furthermore, technical debt has also drawn a lot of attention because it is pricey (Besker et al., 2017; Tom et al., 2013). First, because it is a huge waste in software development time: in average, 36% of all software development time is wasted due of technical debt interest repayment. Second, in 2010, the global technical debt bill was approximately $US500 billion, which may have doubled since. It is therefore critical to consider avoiding and dealing with technical debt as a major issue in software development (Falessi and Kruchten, 2015).

Different issues arise in the technical debt field. These issues have been identified by the main leaders of the field (e.g., Kazman, Kruchten, Ozkaya and Nord) and include estimation issues. In particular, it is very difficult to assess technical debt, since there are different kinds of technical debt and each of them relies on specific measuring techniques. Hence, existing estimation techniques (e.g., techniques implemented by Cast, SonarQube and Jacoco – see reference section) are usually inaccurate, inadequate, insufficient since they mostly focus on requirements, code, tests, usually disregarding key artifacts such as software architecture, build related issues and technologies used by the software at hand. This requires more uniformization but also more dedication to the design and validation of such estimation measures (Falessi and Kruchten, 2015).

Furthermore, one of the most crucial aspects of technical debt but also one of the least explored is its prediction (e.g., Skourletopoulos et al., 2014). Indeed, a software system which has accrued a small amount of technical debt and for which measures have been taken so that it does not accumulate any can still accrue a large amount of technical debt due to unforeseen events. These include changes in supporting technologies trends and even changes in legislation. Such an unexpected accumulation of technical debt in the future might lead a company to bankruptcy. Besides, similar to financial crisis (Investopedia, 2019), technical crisis is a situation in which the value of Tech companies or assets drops rapidly due to the burden of technical debt accumulation. Just as the financial crisis that occurred in 2008 as well as the previous crises, consequences of a technical crisis would be dramatic for the economy. It would drive small, medium and large size companies to financially collapse under the weight of their technical debt, unable to keep the pace with their clients 'needs. This would cause chaotic, harmful and unpredictable technology related disasters (e.g., planes unable to fly, trains unable to operate, medical respirators equipment unable to work properly) due to a shortage in supporting technologies. Foreseeing technical crisis is therefore important to manage it upfront.

In addition, all the existing estimation and prediction approaches are ad hoc approaches that heavily depend on the analyzed systems' languages and platforms. This hinders their reproducibility when applied to different systems' languages and platforms.

To address these shortcomings, it is mandatory that I undertake research focusing on estimation and prediction of technical debt. The main objectives of that research will be the following: 1) improve existing techniques so as to properly evaluate technical debt; 2) determine the extent to which the use of prediction techniques to foresee and therefore avoid accruing technical debt in the future could help companies save money and avoid a potential bankruptcy.

I have broken down the methodology into six phases contributing respectively to the fulfillment of each of the two objectives. I will hire two postdocs and two graduate students to contribute to their realization.

**Phase 1 (April 2019 - August 2019)**: systematically review existing work on technical debt estimation and prediction (e.g., Skourletopoulos et al., 2014) based on the guidelines proposed by Kitchenham (Kitchenham, 2004). This will allow identifying the different artifacts and practices used when estimating and predicting technical debt, as well the literature limits that the proposed research intends to address.

**Phase 2 (September 2019 - December 2019)**: contact companies all around the world to survey technical debt evolution in regard of companies' software artifacts (e.g., code, architecture, commits, comments, requirements, build related issues) as well as technologies (e.g., services, programming languages, databases) used over the past 25 years. This will allow collecting industrial historical data showing the evolution of the technical debt in industrial systems.

**Phase 3 (January 2020 - September 2020)**: rely on the output of the first phase to propose an approach that takes into account all the possible software artifacts and technologies including the ones that are usually neglected by the literature (e.g., architecture, build related issues, technologies at hand) to propose a mathematical calculation model that estimates technical debt using the input currency. This calculation model will be derived from estimation techniques used in finance to estimate financial debt (e.g., Strebulaev and Whited, 2012). The use of estimation models to quantify technical debt is recommended by (Falessi and Kruchten, 2015).

**Phase 4 (January 2020 - May 2020)**: propose an approach that relies on predictive modeling (Kuhn and Johnson, 2013; Tantithamthavorn et al., 2016; Branco et al., 2016) to predict technical debt occurrence. We will define our predictive models by applying machine learning, data mining, and statistics techniques, after a careful review of such techniques to choose the best ones and adapt them in our context. We will train our models using the collected historical data. Note that predictive modeling consists in analyzing data to build a model of an unknown function $Y = f(X1, X2,..., Xp)$, based on a training sample relying on examples of this function (Branco et al., 2016). Our focus will mainly be on technology debt since it seems to be the one that most companies are trying to tackle by migrating for instance toward new technologies at an expensive cost.

In this phase and in the subsequent ones, to ensure the independence of the proposed approaches from the platforms and languages used by the industrial systems involved in our analysis, the historical data retrieved from these systems will be represented using the Knowledge Discovery Meta-Model standard (KDM, 2019). The latter is a standard originally developed by the Object Management Group to represent different aspects of existing systems.

**Phase 5 (June 2020 - September 2020)**: propose an approach able to foresee technical crisis. That approach will rely on predictive modeling, using as input the historical information collected from companies. Our focus will mainly be on technology debt as stated above.

**Phase 6 (October 2020 - March 2021)**: Implement the proposed estimation and prediction approaches in an industrial setting, and carry out experiments on the collected industrial data sets to validate the proposed approaches.

**<u>High reward</u>** – Technical debt leads to high cost of maintenance and evolution when left unmanaged. As stated above, in 2010, the global bill of technical debt was $US500 billion, which may have doubled since. The stakes are therefore high surrounding technical debt management through its estimation and prediction. The proposed research is therefore crucial for companies that want to remain competitive regarding their clients' existing needs as well as the new ones that arise, while taking planned decisions regarding their technical debt accumulation. The automation of the outcomes of this research project will therefore provide to companies a tool serving as a basis for information systems analysis, development and evolution. This research has therefore the potential to make a great economic impact by helping companies save millions by foreseeing technical debt accumulation, technical crisis and therefore avoid a potential bankruptcy in the future.

Besides, the proposed research yields a great potential for patents and publications by being a means to obtain key findings on technical debt. These findings will be promoted through: 1) the obtention of patents; and, 2) publications in well-established scientific venues. In particular, each phase of the methodology described above will result in one or several publications. These publications will have a high impact on the scientific community as well as the industry, by describing in details how the knowledge obtained through the proposed research project can be leveraged. To maximize the impact of these outcomes through publications, my future team members and I will disseminate them by:

- Publishing the results of the research in four top ranked scientific journals, namely: Institute of Electrical and Electronics Engineers Transactions on software engineering, Empirical Software Engineering, Association for Computing Machinery Computer Surveys, and Association for Computing Machinery Transactions on Software Engineering and Methodology.
- Attending and publishing the outcomes of the research in four top ranked scientific conferences, namely: the Technical Debt conference, International Conference on Software Engineering, International Conference on Software Maintenance and Evolution, and International Conference on Software Analysis, Evolution and Reengineering.
- Making available the published research to the general public through online research repositories such as ResearchGate, Google Scholar and Academia.

- Making available the research project and its findings to the general public through a dedicated website where the tool automating the proposed approaches will be downloaded together with its demo and its user manual.

**High risk** – The research project is a high risk project since it defies the current research paradigms by:

1. combining techniques coming from four different fields (i.e. machine learning, data mining, software engineering and finance) belonging to two disciplines (i.e. economics and computer and information sciences) so as to tackle issues raised on technical debt estimation and prediction;
2. collecting historical data from companies across the world to propose novel approaches solving these issues;
3. relying on a larger range of software artifacts and technologies to estimate and predict technical debt;
4. relying on a standard to represent the data so as to support the reproducibility of the proposed approaches that use these data as input
5. being the basis of future collaborations between researchers specialized in economics and those specialized in computer and information sciences
6. being, to the best of our knowledge, the first to coin the expression "technical crisis" in the context of technical debt
7. foreseeing technical crisis in the context of technical debt, which has never been done before.

No related work has ever gone that far, that deep. Besides, to the best of our knowledge, no related work has ever coined the term technical crisis as we do in this proposal, even though the likelihood that it happens is quite high. The Year 2000 bug[1] problem for instance may have led to a technical crisis if it had not been foreseen and handled properly. The proposed research will therefore bring new perspectives to the resolution of the estimation and prediction issues in technical debt.

**Interdisciplinarity** – The proposed research is interdisciplinary since, as Figure 1 depicts, it includes elements from at least two different group-level classifications based on the Canadian Research and Development Classification. The corresponding groups are: 1) Computer and information sciences; and 2) Economics and business administration.

| Code ↑ | Group (Discipline) | Class | Field |
|---|---|---|---|
| RDF1020104 | Computer and information sciences | Artificial intelligence (AI) | Machine learning |
| RDF1020110 | Computer and information sciences | Artificial intelligence (AI) | Data mining |
| RDF1020303 | Computer and information sciences | Programming language and software engineering | Software domains (including operating systems, software infrastructure, etc.) |
| RDF5020202 | Economics and business administration | Business administration and policy studies | Finance |

*Figure 1 Fields of research*

The reliance on an interdisciplinary approach is needed because technical debt is a metaphor that has been borrowed from the finance field. It therefore transposes concepts from the economics discipline to the computer and information sciences discipline. Hence, adapting successful techniques from economics is required to solve some of the limitations found in the literature.

**Equity, Diversity and Inclusion** – the concrete measures that will ensure equity, diversity and inclusion are grouped according to the three key areas described below:

---

[1] https://www.quora.com/What-was-the-Y2K-bug

- **Team composition and training activities**: to make sure that equity, diversity and inclusion principles are key considerations in the composition and management of the research group and training activities, I will welcome and encourage applications from equity-seeking groups (women, Indigenous peoples, members of visible minorities, and persons with disabilities). I will also incorporate the university's Employment Equity objectives and plans in training activities.
- **Recruitment processes**: to make sure that the recruitment of additional team members (faculty, postdoctoral fellows, graduate students, etc.) is open and transparent and aligned with best practices (e.g. minimizing barriers and mitigating against unconscious bias), I will publish the recruitment ads on websites and platforms that are accessible to everyone. These include: 1) the university website; and 2) the software engineering mailing list that is free and accessible to all the software engineering community and is intended to the dissemination of time-sensitive information relevant to the field of software engineering research.
- **Inclusion**: To ensure team members from underrepresented groups are supported and integrated into my future team, I will incorporate the university Equity, Diversity and Inclusion Policy[2] into every single research activity carried out by the team members. I will also offer mentorship to team members coming from underrepresented groups.

## Budget justification:

I will use the allocated fund to pay two postdocs and two master's students who will contribute to the realization of the research project, and who will work under my supervision. Their respective research activities are described below. I will also use the allocated fund to support conference related expenses. These include: conference registration fees, flight tickets, bus tickets, hotel fees, taxis, meals and visa fees. The annual direct costs associated to the proposed research project are described below.

**Postdocs' research activities:**

- The first postdoc will carry out the phases of the research methodology focusing on the estimation of technical debt.
- The second postdoc will carry out the phases of the research methodology focusing on the prediction of technical debt.

**Master's students research activities:**

- The first master's student will assist the first postdoc by carrying out research activities that will help collect and analyze data relevant to the estimation of technical debt.
- The second master's student will assist the second postdoc by carrying out research activities that will help collect and analyze data relevant to the prediction of technical debt.

**Annual costs allocated to year one (2019-2020):**

- Annual salary paid to each postdoc: 36k. Thompson Rivers University will also give, to each postdoc, an annual benefit equal to 23% of his/her annual salary i.e. a benefit equal to $8,280. So, the final annual salary of each postdoc will be $44,280.
- Salary of each master's student involved in the research project for 4 months: 6k.
- Conference expenses for myself, so as to attend two conferences during the year (*): 6k
- Conference expenses for each postdoc attending one conference held in the middle of the year (*) and another one toward the end of the year (**): 5k

**Annual costs allocated to year two (2020 - 2021):**

- Annual salary paid to each postdoc: 36k. Thompson Rivers University will also give, to each postdoc, an annual benefit equal to 23% of his/her annual salary i.e. a benefit equal to $8,280. So, the final annual salary of each postdoc will be $44,280.
- Salary of each master's student involved in the research project for 4 additional months: 6k.

---

[2] https://www.tru.ca/research/research-chairs/equity-diversity-inclusion.html

- Conference expenses for myself, so as to attend two conferences during the year (**): 4k
- Conference expenses for each postdoc attending two conferences during the year (**): 4k
- Conference expenses for each master's student attending one conference during the year (**): 2k.

*(\*) This estimation is an average of the fees computed for a conference held in Occidental Europe (e.g., France, Belgium and Germany) or in Eastern Asia (e.g., China, South Korea)*

*(\*\*) This estimation is an average of the fees computed for a conference held in a neighboring American states (e.g., Washington, Idaho, Montana).*

*Note that all travel and accommodations expenses are estimated based on university rates. Hence, mileage are reimbursed at $0.50 per km (inclusive of all tolls, ferry charges, gas amounts, etc). Meal expenses are reimbursed with original receipts in the following daily maximum allowable amounts: Breakfast $12.00, Lunch $18.00, Dinner $30.00.*

**Total allocated by Thompson Rivers University: $33,120**

**Total requested from the New Frontiers in Research Fund: $200,000.**

# Acknowledgment

I would like to thank Professor Timothy C. Lethbridge and Anita Sharma for reviewing an early draft of this proposal.

# Literature references